\documentclass[prl,twocolumn,showpacs,amsmath,amssymb]{revtex4}
\usepackage{dcolumn}
\usepackage{bm}
\usepackage{graphicx}
\usepackage{color}
\DeclareGraphicsExtensions{.jpg,.pdf,.mps,.png,.eps,.ps,.EPS}

\begin{document}

\def\be{\begin{equation}}
\def\ee{\end{equation}}
\def\bc{\begin{center}} 
\def\ec{\end{center}}
\def\bea{\begin{eqnarray}}
\def\eea{\end{eqnarray}}
\newcommand{\avg}[1]{\langle{#1}\rangle}
\newcommand{\Avg}[1]{\left\langle{#1}\right\rangle}

\title{Statistical Mechanics of the Chinese Restaurant Process: \\ lack
  of self-averaging, anomalous finite-size effects and condensation }
\author{Bruno Bassetti$^1$, Mina Zarei$^2$, Marco
  Cosentino~Lagomarsino$^1$ and Ginestra Bianconi$^3$}

\affiliation{ $^1$ Universit\'a degli Studi di Milano, Dip di Fisica,
  via Celoria 
  16,20133 Milano, Italy \\
$^2$Department of Physics, Isfahan University of
  Technology, Isfahan 84156-83111, Iran \\
  $^3$ Physics Department, Northeastern University, Boston 02115, MA ,USA}

\begin{abstract}
  The Pitman-Yor, or Chinese Restaurant Process, is a stochastic
  process that generates distributions following a power-law with
  exponents lower than two, as found in a numerous physical,
  biological, technological and social systems. We discuss its rich
  behavior with the tools and viewpoint of statistical mechanics.  We
  show that this process invariably gives rise to a condensation,
  i.e. a distribution dominated by a finite number of classes. We also
  evaluate thoroughly the finite-size effects, finding that the lack
  of stationary state and self-averaging of the process creates
  realization-dependent cutoffs and behavior of the distributions with
  no equivalent in other statistical mechanical models.
\end{abstract}

\maketitle

Despite of their extreme behavior, power-law tailed probability
distributions empirically describe the partitioning (i.e. the
organization into sub-classes) of a large variety of physical,
biological, technological and social systems ~\cite{Newman}, and the
degree distributions of many complex
networks~\cite{RMP,Evolution}. The most extreme case are the systems
where the power laws $p_k$ have exponent $\gamma$ between one and two.
Indeed, if $\gamma\in [1,2]$ both the first and the second moment of
the distribution diverge, meaning that the system is so biased that
neither averages nor fluctuations are well-behaved.
This seems to happen for example for words in a text or population of
cities (Zipf's law~\cite{Zipf,Zhang,Newman}), the size of classes of
homolog proteins, the out-degree of transcription
networks~\cite{Kepes}, the frequency of family names \cite{Newman},
and of degree distribution of different social/technological networks
\cite{Hamed}.

While the processes leading to power-law distributions are
diverse~\cite{Newman}, there are only few available models that
capture this behavior. In particular, it is useful to formulate models
that help the understanding of these distributions in the framework of
non-equilibrium growth laws.
The paradigm is the Yule/Simon process~\cite{Yule,Simon}, which
describes an evolving system of a growing number of elements, where
the number of classes grows \emph{linearly} with the elements.
It is based on the general mechanism of the ``Matthew effect'', or
``cumulative advantage'': with time, more populated classes acquire
new elements with higher relative rate.
The Yule/Simon process generates power-law distributions with
exponents $\gamma\in(2,3]$~\cite{Newman}.  Barab\'asi and
Albert~\cite{BA} have shown that a similar mechanism can be used to
generate power-law networks with the same exponents that grow and
evolve through preferential attachment~\cite{BA,RMP,Evolution}.

While in some systems where exponents $\gamma\leq 2$ are observed it
can be argued that preferential attachment is present, this case is
not predicted by the Yule/Simon/Barab\'asi-Albert (YSBA) model.
These distributions are more biased towards highly populated classes,
so that in order to obtain this behavior one has to reweigh the
balance of growth and preferential attachment in favor of the latter,
and in particular consider processes where the number of classes grows
\emph{sub-linearly} with the number of elements.

The process we consider here, called Pitman-Yor, or Chinese Restaurant
Process (CRP) \cite{Yor,Pitman}, has exactly this property, which
makes it reproduce power-laws with $\gamma\in (1,2]$. 
It is commonly used in the mathematics literature, but relatively
disregarded by the physics community, and in particular unexplored
using the tools of statistical mechanics. 
It is defined as a discrete-time stochastic process generating a
partition of a number elements in classes, such that each element
belongs to a given class.  In probability, the CRP is used for example
as a prior in nonparametric Bayesian methods and its has been applied
to multiple problems ranging from modeling texts to genetics and
functional genomics \cite{Blei,Markov,Microarray,Haplotypes,Ewens}. It
also maps to a non self-averaging stick-breaking
process~\cite{Pitman,Derrida}.  Recently, we observed that CRP-like
processes model well the evolution protein domain families
\cite{domains} and reproduce the scaling laws found by genomics
methods, which adds a strong motivation to explore them.

The mathematical characterization of the CRP has been carried out
\cite{Pitman} with special attention to the asymptotic of the process
at large times $T\rightarrow \infty$.
In this Letter, we characterize this process with the tools of
statistical mechanics. 
First, we argue that the CRP always exhibits a \emph{condensation}
phenomenon \cite{Burda,Bose,review,Noh,Majumbdar,Critical,Godreche}
with few classes dominating the total population. We relate the
condensation observed in the CRP process to other known mechanisms
taking place in the well-studied phenomenology of the Zero-Range
Process~\cite{review}, or in network models~\cite{Bose}.
Second, we present a calculation that shows how the process behaves
for large but finite times $T$, finding anomalous finite-size
corrections to the asymptotic formulas.
Unlike the YSBA model~\cite{Redner}, the lack of self-averaging of the
CRP determines for certain parameter values a nontrivial and
realization-dependent finite-size behavior which is our main 
finding.  Thus, the CRP fills two important gaps in the fundamental
statistical mechanical understanding of non-self averaging phenomena,
power-law distributions, and condensed states.

\vskip 0.2 cm

{\it General Considerations.}  At each time $T$, the CRP generates a
partition over integers $\{1,2,\ldots T\}$ into different classes.
Differently from other models \cite{Bose,review}, the number of
classes $N(T)$ is a stochastic variable depending on the realization.

The process is anecdotally a problem of customers entering a Chinese
Restaurant with table-sharing, where the number of tables and guests
per table are unbounded.  Assuming that in the restaurant there are
$T$ customers sitting at $N(T)$ tables with $k_i$ customers in each
table $i=1,2,\ldots N(T)$.  At time $T$ a new customer enters the
restaurant and either sits at table $i=1,\ldots, N(T)$ with
probability $p_i$ (in this case $N(T+1)=N(T)$), or chooses a new table
with probability $p_{N(T)+1}$ (in this case $N(T+1)=N(T)+1$).  The
probability $p_i$ and $p_{N(T)+1}$ in the CRP are given by
\bea
p_i=\frac{k_i-\alpha}{T+\theta} &\ \ & p_{N(T)+1}=\frac{\alpha N(T)+\theta}{T+\theta}, 
\label{pp}
\eea 
where $\theta>0$ and $\alpha\in[0,1)$.

As in the YSBA model, the CRP includes growth of elements $T$ and
classes $N$, and a preferential attachment principle, because more
populated tables are more likely to acquire new customers.
However, in the CRP, growth of classes is not constrained to
preferential attachment of new class members, but these two processes
are decoupled, as witnessed by the fact that the probability to add a
new table is not constant.  This probability decays with the number of
guests $T$, increasing the weight of ``hub classes'', which is the
essential ingredient to reproduce power-law distributions with
exponent lower than two.  While models producing power-laws with
$\gamma\in (1,2]$ exist~\cite{Referee1}, usually they lack the
flexibility of the CRP in modulating this weight.

Rephrasing the YSBA model in terms of customers in a restaurant,
the probabilities $p_i^{S}$ to sit at a non-empty table $i$ at time
$T$ and the probability $p_{N(T)+1}^{S}$ to sit at a new table at time
$T$ is $p_i^{S}=(1-\varepsilon){k_i}/{T},$
$p_{N(T)+1}^{S}=\varepsilon$ where $k_i$ is the occupation number of
table $i=1,\ldots,N(T)$. Note that this means that new tables are
added with statistically independent moves, while in the CRP the
addition of a new table is statistically \emph{dependent} on the
configuration of the partition~\cite{domains}. 

The CRP has been studied extensively in the mathematical literature
\cite{Pitman}.  The occupation distribution in the limit $T\rightarrow
\infty$ and $k$ finite and fixed is
$F_{\infty}(k)=\frac{\Gamma(k-\alpha)}{\Gamma(k+1)\Gamma(1-\alpha)}$.
Furthermore, the statistics of the number of tables $N(T)$ has been
characterized.  The average value of the number of tables $\avg{N(T)}$
at time $T$ is given by \cite{Pitman}
\bea
\avg{N(T)}\simeq\left\{\begin{array}{lcr}
    \frac{\Gamma(\theta+1)}{\alpha\Gamma(\theta+\alpha)}T^{\alpha}&
    \mbox{ for } & \alpha>0\nonumber \\ \theta \log(T+\theta)
    &\mbox{for }& \alpha=0
\end{array}\right.
\eea 
In the limit of large $T$, the full probability distribution for
$N(T)$ ${\cal P}(N(T))$ is known \cite{Pitman}, when $\alpha=0$, to be a
Gaussian of mean $m$ and standard deviation $\sigma^2$ with
$m=\sigma^2=\theta\log(T)$. In the case $\alpha>0$ , instead, the
variable $s=N(T)/T^{\alpha}$ asymptotically in time follow the
Mittag-Leffler distribution $g_{\alpha,\theta}(s)$~\cite{Pitman}.
This point is particularly interesting \cite{Pitman,Sibuya} because in
the asymptotic limit, the Mittag-Leffler distribution has finite
fluctuations, implying that the number of tables in the Pitman-Yor
process with $\alpha>0$ is a non self-averaging quantity.

{\it The CRP is always in a condensed state.}  We now provide an
argument comparing the phenomenology of the CRP to models exhibiting
condensation phenomena.
Extending the validity of the asymptotic formula $F_{\infty}(k)$ for
all values of $k$, we can estimate the occupation of the maximally
occupied table in the CRP. We observe that this table has always
occupation $k_{max}={\cal O}(T)$.  In fact, since $F_{\infty}(k)\sim
k^{-\alpha-1}$, we can evaluate the occupation $k_{max} $ of the
maximally occupied table by imposing the defining condition that that
the fraction of tables with $k>k_{max}$ must be of the order of $1/N$,
i.e.
\begin{equation}
\sum_{k>k_{max}}F_{\infty}(k)\simeq\frac{1}{N}.
\end{equation} 
Since in the Chinese restaurant process $N={\cal O }(T^{\alpha})$ if
$\alpha>0$, and $N={\cal O}(\log(T))$ if $\alpha=0$, in both cases
this estimate indicates that the maximally occupied table has a finite
occupation $k_{max}={\cal O}(T)$.

When the maximal occupation of a class is of the same order of the
total number of elements in the partition, one says that the
distribution is in a "condensed" phase.  Reference models studied in
the statistical physics community are the Zero-Range-Process (ZRP) and
the Bose-Einstein condensation of networks (BECN) \cite{Bose}. In the
BECN the condensation occurs on a single
special node, for power-law degree distributions with exponent  $\gamma=2$  as a consequence of the heterogeneity of the classes.
In the ZRP, particles hop on 1-D lattice sites according to prescribed
laws \cite{review}, generating partitions of elements into classes,
i.e.  clusters of particles, with power-law behavior and exponent
$\gamma$. Depending on the dynamics and particle density, a condensation can
occur in the ZRP, where one class becomes occupied by a finite
fraction of elements.  

It is instructive to illustrate the main differences between the
condensation phenomena occurring in the ZRP and in the CRP: First, in
the ZRP the exponent $\gamma$ of the distribution can be larger or
smaller than $2$, but the condensation occurs only if $\gamma>2$,
while in the CRP a condensation always occurs and the distribution of
the partition decays with an exponent $\gamma=1+\alpha<2$.  Second, in the
$\gamma>2$ ZRP, the condensation transition is driven by the density
of particles $\rho$: If $\rho>\rho^{\star}$ there is a condensation,
if $\rho<\rho^{\star}$ there is no condensation, and the condensate
appears in order to balance between the imposed finite value of $\rho$
and the natural average value of the power-law. Conversely, in the CRP
the mean density of elements always \emph{diverges}, which, in the
large $T$ limit imposes the existence of classes with a finite
fraction of the total number of elements.
Thus, a relevant difference between the ZRP (and BECN) and the CRP is
that in the CRP there is a degenerate distribution but no phase
transition.  This situation closely resembles the so called
``pseudo-condensation'' found \cite{Evans2} where the condensation is
characterized in a ZRP with non-extensive number of classes.  However
while in that case the scaling of the number of classes with the number
of elements is chosen \emph{ad hoc}, in the CRP this scaling is a
natural outcome the process.

\vskip 0.2 cm

{\it Conditioned Path Integral of the CRP and anomalous finite-size
  effects.}  
At finite-sizes the $\alpha>0$ CRP shows an intriguing 
phenomenology, where the trend of individual realizations determines
their distribution.  This is visible from the finite-size scaling of
the distribution $F(k,T)$.  We thus study $F(k,N,T)$, with the
additional condition of \emph{fixed} number of tables $N$.

The probability of a partition of $T$ elements is the probability
distribution at $T-1$ times the probability of an event at time $T$.
Therefore the probability $P(\{k_i\})$ of a process from time $T=1$ to
time $T$, giving rise to an occupation of $N$ tables $i=\{1,2,\ldots
N\}$, each one occupied by $k_i$ individuals, is given by the product
of the probabilities $(\ref{pp})$ for each subsequent event. In
particular this probability can be written as
\bea 
\hspace*{-2mm} P({\{k_i\}})=C_{N,T}\left(\prod_{i} \frac{
    \Gamma(k_i-\alpha)}{\Gamma(1-\alpha)}\right)\delta_{T,\sum_i k_i} \ ,
\label{Piki}
\eea 
where $\delta$ is the Kronecker delta fixing the total number of
customers and the constant $C_{N,T}$ is given by $
C_{N,T}={\alpha^N}{\Gamma(N+\theta/\alpha)\Gamma(\theta)}/[{\Gamma(\theta/\alpha)\Gamma(T+\theta)]}$.
Most notably, the probability $P(\{k_i\})$ of a process giving rise to
the occupation numbers $\{k_i\}$, Eq. $(\ref{Piki})$ is independent on
the history of the process.  In this case $P(\{k_i\})$ is called a
distribution of exchangeable random variables \cite{Pitman}. Moreover,
since $P(\{k_i\})$ takes a factorizable form, this probability
distribution is also referred to as a Gibbs measure \cite{Pitman}.

We can construct a conditioned path integral of this process by
summing over all the histories keeping $T$ and $N$ constant. Since the
events in the CRP are exchangeable~\cite{Pitman}, i.e. the probability
is invariant for any permutation of the set of class indexes, we can
sum over the histories in which the partition $\{k_i\}$ is generated
in random order. To account for the number of these histories we introduce  the multinomial
prefactor $T!/(\prod_i k_i! \, N!)$. This leads to the following
expression for the partition function $Z_{N,T}$,
%
\be
\hspace*{-1mm}Z_{N,T}= \frac{1}{N!}\sum_{\{k_i\}_{i=1,\ldots N}}\frac{T!}{\prod_i
  k_i!}P({\{k_i\}}) \ .
\label{ZNT}
\ee
%
Similarly, the probability $F(k,N,T)$ that in a process studied at
time $T$ when $N$ tables are full, a random table is occupied by $k$
guests reads
\bea
\hspace*{-4mm}
F(k,N,T)=  \frac{1}{Z_{N,T}}\sum_{\{k_i\}_{i=1,\ldots
    N}}\frac{T!}{N!\prod_i k_i!}\delta_{k_1,k}P({\{k_i\}}) .
\eea
Or equivalently,
\be F(k,N,T)=\frac{Z_{N,T-k}}{Z_{N,T}}
\frac{\Gamma(k-\alpha)}{\Gamma(k+1)\Gamma(1-\alpha)}
\label{f1}
\ee 

Roughly, the ratio appearing at the r.h.s. in this equation is related
to the power-law behavior, while the rest gives the finite-size
corrections.  The function $Z_{N,T}$ can be evaluated, for large $T$,
with a saddle-point approximation of the integral
\be \hspace*{-5mm}Z_{N,T}=\int \frac{d\omega}{2\pi} e^{i\omega
  T}\left(\sum_{k= 1}^T
  \frac{\Gamma(k-\alpha)}{\Gamma(1-\alpha)\Gamma(k+1)}e^{-i\omega
    k}\right)^N \ ,
\ee
where the integration over $\omega$ comes from the Fourier
representation of the Kronecker delta of Eq.~\ref{Piki}, and the saddle
point $i\omega^{\star}=\hat{\omega}:=\chi/T$ satisfies the equation
\be
\frac{T}{N}=\frac{H_1(T,\alpha,\chi/T)}{H_0(T,\alpha,\chi/T)},
\label{sp}
\ee 
where 
\be H_n(T,\alpha,\omega)=\sum_{m=1}^T
m^n\frac{\Gamma(m-\alpha)}{\Gamma(1-\alpha)\Gamma(m+1)}e^{-\omega m} \
\ .  \ee

For $\alpha>0$, the solution to the saddle-point equation
$(\ref{sp})$, $\chi=\chi(N,T,\alpha)$ in the limit $T\rightarrow
\infty$ depends on $N,T$ only through the realization-specific
variable $s=N/T^{\alpha}$. In order to show this, we first observe
that the scaling of $H_n(T,\alpha,\chi/T)$ with $T$, can be studied by
approximating integral to sums in their definition.  Secondly, we show
that the functions $H_n(T,\alpha,\chi/T)$ can be expressed as
\bea H_n(T,\alpha,\chi/T)&\simeq &\frac{1}{\Gamma(1-\alpha)}
\frac{[T^{n-\alpha}-1]} {(n-\alpha)}
\nonumber \\
&&\hspace*{-30mm}+\frac{1}{\Gamma(1-\alpha)} T^{n-\alpha}\int_0^1 dt\:
\frac{1}{t^{\alpha+1-n}} \: (e^{-\chi t} -1) \ ,  \eea 
where we have added and subtracted a term of the type
$H_n(T,\alpha,0)$.  Consequently, for large $T$,
$H_1(T,\alpha,\chi/T)\rightarrow T^{1-\alpha}h_1(\alpha,\chi) $ while
$H_0(T,\alpha,\chi/T)\rightarrow h_0(\alpha)$. Inserting these relations
in the saddle point-equation $(\ref{sp})$, and taking $N=sT^{\alpha}$
we obtain
\be \frac{1}{s}=\frac{h_1(\alpha,\chi)}{h_0(\alpha)} \ee
proving that $\chi=\chi(s,\alpha)$ in the large $T$ limit for
$\alpha>0$.

The final expression for $Z_{N,T}$ is therefore given by 
\be
Z_{N,T}\simeq \frac{C_{N,T}}{N!} e^{\chi(s,\alpha)}\frac{e^{N
    \log(H_0(\alpha,\chi(s,\alpha)/T, T))}}{\sqrt{N J(T,\alpha,\chi/T)}}
    \label{znt}
\ee
where we evaluated  the saddle point   up to
the second order, and the function $J(T,\alpha,\chi/T)$ is given by
\be
J(T,\alpha,\chi/T)=\left.\frac{\partial^2
  \log(H_0(T,\alpha,\omega)}{\partial \omega^2}\right|_{\omega=\chi/T}.
\ee
A similar procedure applies to the evaluation of $Z_{N,T-k}$, with
$\chi'=\omega'/(T-k)$ satisfying the saddle point equation \be
\frac{T-k}{N}=\frac{H_1(T-k,\alpha,\chi'/(T-k))}{H_0(T-k,\alpha,\chi'/(T-k))}.
\label{sp2}
\ee 
Following arguments similar to the one provided for the scaling of
$\chi$, we can show that $\chi'=\chi'(k,N,T,\alpha)$ at the saddle
point. Equation $(\ref{sp2})$ depends on $k,N,T$ only through the
variables $s=N/T^{\alpha}$ and $k/T$, i.e.
$\chi'=\chi'(s,k/T,\alpha)$. Following a similar reasoning to he one
we adopted for proving that the function $Z_{N,T}$ in
Eq. $(\ref{znt})$ depends exclusively on the parameters $s,\alpha$, it
is possible to show that the function $Z_{N,T-k}$ only depends on
$s,k/T, \alpha$, i.e. $Z_{N,T-k}=\phi(s,k/T,\alpha)$.

\begin{figure}
\includegraphics[width=65mm,height=55mm]{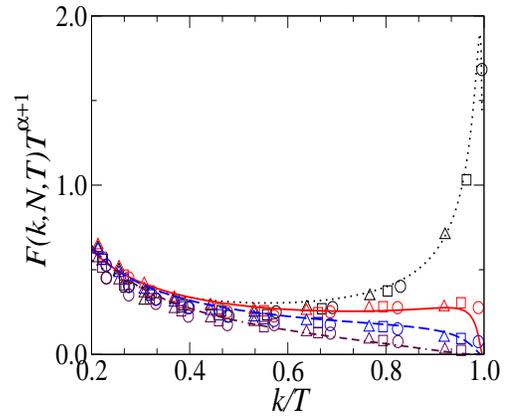}
\caption{(Color online) Rescaled distribution of the occupation
  numbers in the Chinese Restaurant Process with $\alpha=0.1$ and
  $\theta=1$ at different times $T$ and number of tables
  $N=sT^{\alpha}$. We report the log-binned distribution $F(k,N,T)$
  for $s=N/T^{\alpha}$ with $s=3.5$ (black symbols, dotted line) $s=5$
  (red symbols, solid line) $s=7.5$ (blue symbols, dashed line) $s=10$
  (brown symbols, dashed-dot line). The rescaled data are shown for
  processes with $T=2500$ (triangles),$T=5000$ (squares), $T=10^4$
  (circles). The solid line show the analytical solutions calculated
  by solving the saddle point equations $(\ref{sp})$, $(\ref{sp2})$
  for $T=2500$.}.
\label{scaling.fig}
\end{figure}

Therefore, taking in $(\ref{f1})$ the large $k, T$ limit with
$k/T={\cal O}(1)$, $F(k,N,T)$ for $\alpha>0$ satisfies the scaling
relation
\begin{eqnarray}
T^{\alpha+1}F(k,N,T)&=&T^{\alpha+1}\frac{Z_{N,T-k}}{Z_{N,T}}
\frac{\Gamma(k-\alpha)}{\Gamma(k+1)\Gamma(1-\alpha)}\nonumber \\
&=&\left(\frac{T}{k}\right)^{1+\alpha} q(k/T,s=N/T^{\alpha},\alpha) \ ,
\label{scaling.eq}
\end{eqnarray}
where the function $q$ (containing $\chi'$, $\chi$, and the second
order corrections) represents the finite-size corrections to the
asymptotic behavior.  These findings shed light on the absence of
self-averaging in the process.  At any given time $T$ the process will
have finite fluctuations, persistent also in the limit of $T \to
\infty$.  These fluctuations depend on the non stationarity of the
process, and on the non self-averaging value of the number of classes
$N$.  Therefore the process, if conditioned on the number of classes
$N$, shows fluctuations that go to zero as $T \to \infty$.  Figure
$\ref{scaling.fig}$ compares simulations with the analytical
predictions of Eq. (\ref{f1}). The figure shows that the finite-size
correction to the power-law tail $\sim 1/k^{1+\alpha}$, for some $s$
and large $k/T$ may \emph{increase} its value, giving rise to an
anomalous ``bump'' in the distribution.
On the other hand, this local maximum never develops into a
concentrated ``condensate'', and for $k/T \to 1$, for any $s$, the
cutoff $q$ always dampens $F$. 

In conclusion, we have presented a statistical mechanics study of the
Chinese Restaurant Process, which generates power-law distributions
with exponents $\gamma \in(1,2]$ by nonequilibrium growth, a
condensation phenomena, absence of selfaverging, and anomalous
finite-size effects.
We believe that this rich process will be of importance for future
developments of the field where these trends occur, i.e.  biological
evolution, complex systems, spin-glasses and nonequilibrium phenomena.

G.B acknowledges support from the IST STREP GENNETEC contract number
034952.

\end{document}